\documentclass[fleqn,usenatbib]{mnras}

\usepackage{newtxtext,newtxmath}

\usepackage[T1]{fontenc}
\usepackage[justification=centering]{caption}
\DeclareRobustCommand{\VAN}[3]{#2}
\let\VANthebibliography\thebibliography
\def\thebibliography{\DeclareRobustCommand{\VAN}[3]{##3}\VANthebibliography}

\usepackage{graphicx}	
\usepackage{amsmath}	
\usepackage{amsfonts,bm}
\usepackage{ulem} 

\renewcommand{\vec}[1]{\bm{#1}}

\def\inp{\langle\delta h|\delta h\rangle}

\title[Detecting lensed SMBHBs by LISA]{A Higher Probability of Detecting Lensed Supermassive Black Hole Binaries by LISA}

\author[Gao et al.]{
Zucheng Gao, $^{1,2}$\thanks{E-mail: phy15gzc@pku.edu.cn}
Xian Chen, $^{1,3}$\thanks{E-mail: xian.chen@pku.edu.cn}
Yi-Ming Hu, $^{4}$
Jian-Dong Zhang $^{4}$
and Shun-Jia Huang $^{4}$
\\
$^{1}$Astronomy Department, School of Physics, Peking University, Beijing 100871, P. R. China\\
$^{2}$Institute of Astronomy, University of Cambridge, Madingley Road, Cambridge CB3 0HA, UK\\
$^{3}$Kavli Institute for Astronomy and Astrophysics at Peking University, Beijing 100871, P. R. China\\
$^{4}$TianQin Research Center for Gravitational Physics and School of Physics and Astronomy, Sun Yat-sen University (Zhuhai Campus), Zhuhai 519082, P. R. China
}

\date{Accepted XXX. Received YYY; in original form ZZZ}

\pubyear{2021}

\begin{document}
\label{firstpage}
\pagerange{\pageref{firstpage}--\pageref{lastpage}}
\maketitle

\begin{abstract}
Gravitational lensing of gravitational waves (GWs) is a powerful probe of the
matter distribution in the universe. Here we revisit the wave-optics effects
induced by dark matter (DM) halos on the GW signals of merging massive black
hole binaries (MBHBs), and we study the possibility of discerning these effects using the Laser
Interferometer Space Antenna (LISA).  In particular, we include the halos in
the low-mass range of $\rm 10^5-10^8\, M_\odot$ since they are the most
numerous according to the cold DM model. We simulate the lensed signals
corresponding to a wide range of impact parameters, and we find distinguishable
deviation from the standard best-fit GW templates even when the impact
parameter is as large $y\simeq50$.  Consequently, we estimate that over
$(0.1-1.6)\%$ of the MBHBs in the mass range of $\rm 10^{5.0}-10^{6.5}\,
M_\odot$ and the redshift range of $4-10$ should show detectable wave-optics
effects.  This probability is one order of magnitude higher than that derived
in previous works.  The uncertainty comes mainly from the mass function of the
DM halos.  Not detecting any signal during the LISA mission would imply that DM
halos with $\rm 10^5-10^8\, M_\odot$ are less numerous than what the cold DM model predicts.
\end{abstract}




\section{Introduction}
\label{sec:intro}

Gravitational lensing events are unique probes of the distribution of matter
in the universe \citep{schneider92}.  Like light, gravitational waves (GWs)
could also be lensed by intervening matter
\citep{lawrence71,cyranski74,sonnabend79,markovic93}.  The prospect of 
detecting the lensing of GWs is promising given the
increasing number of GW events discovered in the recent years by the Laser
Interferometer Gravitational-wave Observatory (LIGO) and the Virgo detectors
\citep{Abbott_2016,Abbott_2019a,abbott20catalog}.

One major difference between GW and light is that the former usually has a much
longer wavelength.  For example, LIGO/Virgo are sensitive to the GWs with a
wavelength of ${\cal O}(10^4)~{\rm km}$. It is comparable to or longer than the
characteristic sizes of many astrophysical objects, such as stars or
intermediate-massive black holes (IMBHs). If lensed by these objects, the GWs
in the LIGO/Virgo band would behave like light in the wave-optics limit
\citep{ohanian74,bontz81,nakamura98,1999PThPS.133..137N}.  In this case, wave
diffraction would modify the amplitude and phase of the GWs, producing a
characteristic ``beating pattern'' in the frequency domain of the waveform
\citep{takahashi03}. In addition, the wave-optics effect could also smear
the plane of GW polarization \citep{cusin20} and produce beat patterns in the
time-domain waveform \citep{yamamoto05,hou20}. These effects, in principle,
could allow LIGO/Virgo to detect massive stars, IMBHs, and the dense cores of
globular clusters and dark-matter (DM) halos
\citep{moylan08,cao14,takahashi17,christian18,2018PhRvD..98j4029D,diego19,jung19,2019ApJ...875..139L,meena20,oguri20,mishra21,Wang:2021}.
However, so far no strong evidence of lensing effects has been officially reported by LIGO/Virgo
\citep{hannuksela19}, suggesting that the lensing probability is relatively
low. 

The Laser Interferometer Space Antenna (LISA) is a future space-based mission
aiming at detecting the GWs in the milli-Hertz (mHz) band
\citep{Amaro-Seoane_2017}. One of its major targets is the merger of two
massive black holes (MBHs), preferentially in the mass range of
$10^4\sim10^7\,M_\odot$. Because of the superb sensitivity,  LISA could detect
MBH mergers up to a redshift of $20$ with a signal-to-noise ratio (SNR) as high
as $10^2-10^3$ \citep{Amaro-Seoane_2017}.  Such a high redshift suggests that
gravitational  lensing by the large-scale structure is no longer negligible for
LISA \citep{takahashi06,yoo07}. The long wavelength and high SNR also indicate
that the diffraction effects in the wave-optics limit, which is relatively weak
for LIGO/Virgo sources, may become significant for LISA.  

For LISA, the lenses which produce the diffraction effects are mainly low-mass
dark-matter (DM) halos, as well as the subhalos in massive main halos
\citep{takahashi03,takahashi04}.  \citet{takahashi03} considered the DM
halos in the mass range of $10^9\sim10^{12}M_\odot$ and estimated that the
lensing probability for each MBH merger in the LISA band is about
$10^{-4}\sim10^{-3}$. However, the cold DM (CDM) model predicts that the most
abundant halos are those in the mass range of $10^6\sim10^9M_\odot$
\citep[e.g.][]{COORAY_2002,Han_2016}.  These small halos are not included in
the calculation of \citet{takahashi03}.  Moreover, \citet{takahashi03}
imposed a maximum impact parameter $y=3$ for the sources, to ensure a strong,
detectable diffraction effect. This criterion, however, may be too strict. The
high SNR of LISA sources would enable us to detect weak lensing signals even
when the impact parameter is larger. 
Since the lensing probability increases with $y^2$,
previous studies could have significantly underestimated
the detection rate of lensed signals.

To elucidate the real detection rate, we revisit the 
signals of the lensed MBHs in the LISA band. We pay special attention
to the events with large impact parameters and we employ the matched-filtering
technique to search for weak wave-optics effects in the signal.
We also improve the
lens model by including the halos and subhalos at the lower mass end.
We remark that although the calculation is based on LISA, the conclusion
also applies to 
space-borne GW missions with slightly different frequency coverage, 
such as TianQin \citep{Mei:2020lrl}.
The paper is organized as follows.  In
Section~\ref{sec:methods}, we describe our method, including the calculation of
the lensing signal in the wave-optics limit, the mass function of halos and
subhalos, the probability of lensing, as well as the matched-filtering
technique.  Then in Section~\ref{sec:diff} we quantify the difference between
the lensed and unlensed signals and derive a criterion for detecting the
diffraction effect. Using this criterion, we estimate the lensing probability
for the MBH mergers in the LISA band in Section~\ref{subsec:LP}. We discuss the
importance of DM models in Section~\ref{sec:DMmodel} and conclude in
Section~\ref{sec:conclusion}.  Throughout the paper, we assume a flat
$\mathrm{\Lambda CDM}$ cosmology with the parameters $\Omega_m=0.27$,
$\Omega_\Lambda=0.73$, $H_0=72\rm kms^{-1}Mpc^{-1}$, $\sigma_8=0.9$.
Therefore, in our work $h:=H_0/(100~\rm kms^{-1}~Mpc^{-1})=0.72$.

\section{Method} \label{sec:methods}
\subsection{Lensing Model} \label{subsec:lensing_model}

For simplicity, we assume a singular isothermal sphere (SIS) profile for our DM halos and subhalos.  In
this case, we can follow \citet{takahashi03} to model the diffraction of
GWs.  Using a more realistic Navarro-Frenk-White (NFW) profile usually leads to a slightly smaller
magnification factor \citep{takahashi03,Choi:2021}.

\begin{figure}
    \centering
    \includegraphics[width=\linewidth]{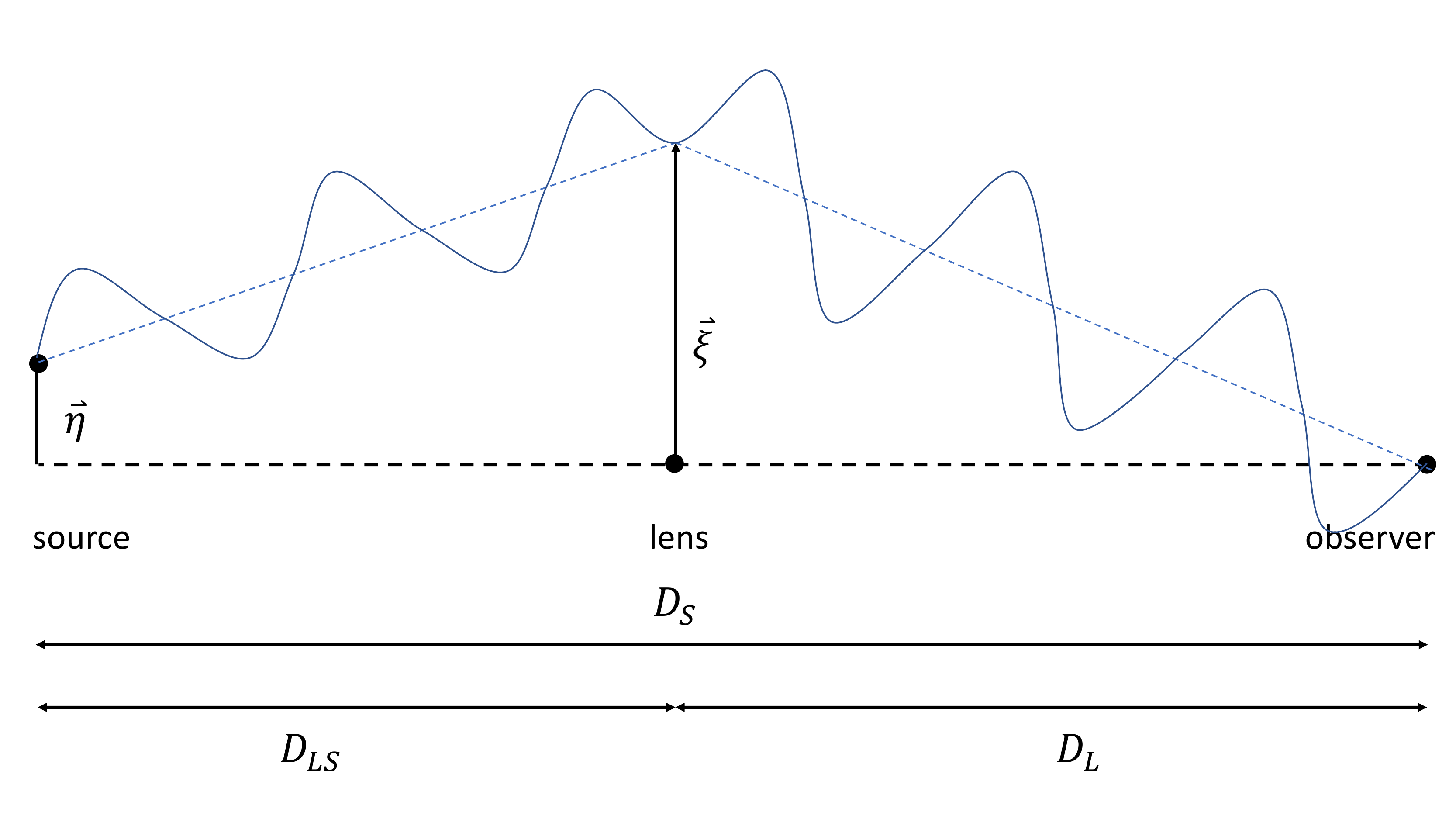}
    \caption{Physical picture of GW lensing. The vectors $\vec{\eta}$ and $\vec{\xi}$ 
denote the two positions on the source plane and on the lens plane. 
For illustrative purposes the two vectors are aligned but in principle
	they are not.
The angular diameter distances 
 $D_L$, $D_S$, and $D_{LS}$ are measured in the rest frame of the observer.}
    \label{Fig1}
\end{figure}

The basic picture of lensing of GWs is illustrated in  Figure~\ref{Fig1}, where
$D_L$, $D_S$, and $D_{LS}$ denote the angular diameter distances to the lens,
to the source, and their difference.  All these quantities are measured in the
frame of the observer.

The magnification factor is defined as
\begin{eqnarray}\label{Eq2}
	F(\omega, \vec{\eta})&:=\frac{\Tilde{\phi}^L_{obs}(\omega,\vec{\eta})}{\Tilde{\phi}_{obs}(\omega,\vec{\eta})},
\end{eqnarray}
where $\Tilde{\phi}^L_{obs}(\omega,\vec{\eta})$ and
$\Tilde{\phi}_{obs}(\omega,\vec{\eta})$ are the lensed and unlensed
gravitational wave amplitudes in the Fourier space, $\omega$ is the observed
angular frequency of the GW, and $\vec{\eta}$ is the position vector on the
source plane (``impact parameter'' hereafter).  Taking the cosmological
redshift into account, the magnification factor can be calculated with
\begin{eqnarray}\label{Eq3}
	F(\omega, \vec{\eta})& = \frac{\omega(1+z_L)D_S}{2\pi i\,D_L D_{LS}}\int d^2\xi \exp[i\omega(1+z_L)t_d(\vec{\xi},\vec{\eta})],
\end{eqnarray}
where $z_L$ is the redshift of the lens and
$t_d$ is the time delay caused by lensing.
The time delay can be computed with
\begin{eqnarray}\label{Eq4}
    t_d(\vec{\xi},\vec{\eta}) = \frac{D_L D_S}{2D_{LS}}\left|\frac{\vec{\xi}}{D_L}-\frac{\vec{\eta}}{D_S}\right|^2-\hat{\psi}(\vec{\xi})+\hat{\phi}_m(\vec{\eta}),
\end{eqnarray}
where $\hat{\phi}_m(\vec{\eta})$ denotes the arrival time of the unlensed GW, which 
is approximately $D_S[1+|\vec{\eta}/D_S|^2/2]/c$, and $\hat{\psi}$ is the
deflection potential which solves the equation
\begin{eqnarray}\label{Eq5}
    \nabla_{\xi}^2\hat{\psi}(\vec{\xi}) = 8\pi \Sigma(\vec{\xi}),
\end{eqnarray}
with $\nabla_{\xi}^2$ the two-dimensional Laplacian and $\Sigma(\vec{\xi})$ the
mass surface density of the lens. 

For the purpose of numerical calculation, we define the dimensionless positions as
\begin{eqnarray}\label{Eq6}
    \vec{x}=\frac{\vec{\xi}}{\xi_0};\, \vec{y}=\frac{D_L}{\xi_0 D_S}\vec{\eta},
\end{eqnarray}
where $\xi_0$ is the Einstein radius. In the SIS model, it can be calculated with $\xi_0 = 4\pi \sigma_v^2 D_L D_{LS}/c^2D_S $,
where $\sigma_v$ is the velocity dispersion of the lens.
The corresponding dimensionless frequency is
\begin{eqnarray}\label{Eq7}
    {w} = 4GM_{L}(1+z_L)\omega/c^3,
\end{eqnarray}
where $M_{L}$ is the so-called ``lens mass'', which is defined as the
mass enclosed by a circle of the Einstein radius in the lens plane.
In the SIS model, we have $M_{L} = 4\pi^2\sigma_v^4D_LD_{LS}/(GD_Sc^2)$
(Appendix~\ref{append:A}). 
Using the above nondimensional quantities,
the time delay can be rewritten as
\begin{eqnarray}\label{Eq8}
	&T&(\vec{x},\vec{y})=\frac{D_LD_{LS}}{D_S}\xi_0^{-2}t_d(\vec{\xi},\vec{\eta})
\nonumber	\\
	&=&\frac{1}{2}|\vec{x}-\vec{y}|^2-\frac{D_LD_{LS}}{D_S}\xi_0^{-2}\hat{\psi}(\vec{\xi})+\frac{D_LD_{LS}}{D_S}\xi_0^{-2}\hat{\phi}_m(\vec{\eta}).
\end{eqnarray}
It follows that the nondimensional amplification factor is
\begin{eqnarray}\label{Eq9}
    F(w,\vec{y})=\frac{w}{2\pi i}\int d^2x\exp[iwT(\vec{x},\vec{y})].
\end{eqnarray}
In the SIS model the last equation can be calculated with
\begin{eqnarray}\label{Eq10}
	F(w, y) &=& -iw e^{iwy^2/2} \int_0^{\infty} dx ~x J_0(wxy)\times\nonumber\\
	&&\exp[iw(\frac{1}{2}x^2-x+\phi_m(y))],
\end{eqnarray}
where $\phi_m(y)=y+1/2$ and $J_0$ is the zeroth-order Bessel function.
The corresponding amplification factor $|F|$ and phase-change factor $\theta_F$ are
\begin{eqnarray}\label{Eq11}
   |F| = \sqrt{FF^*},\,\, \theta_F = -i\ln[F/|F|],
\end{eqnarray}
where $F^*$ is the complex conjugate of $F$.

\begin{figure}
    \centering
    \includegraphics[width=0.45\textwidth]{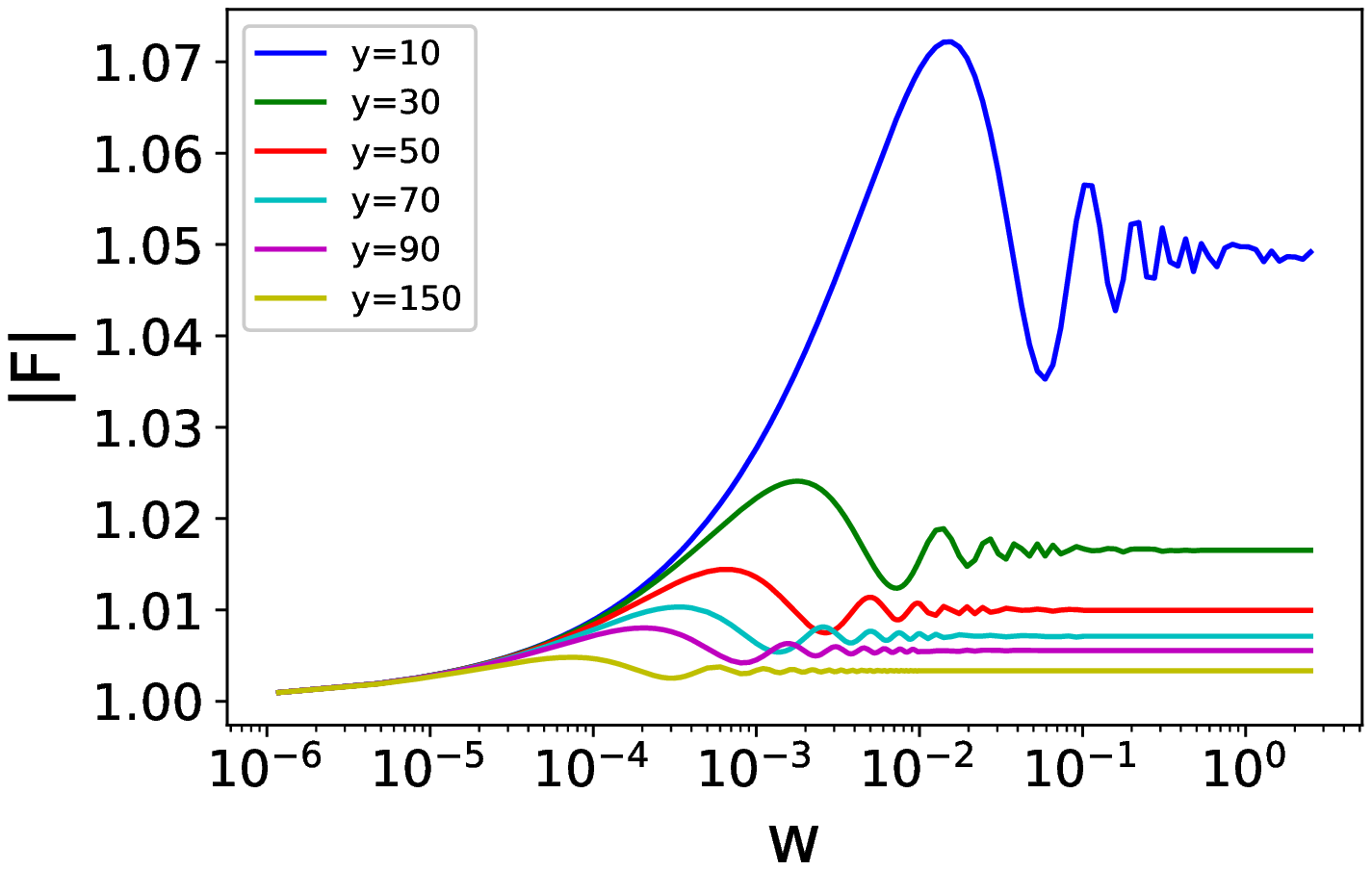} \includegraphics[width=0.45\textwidth]{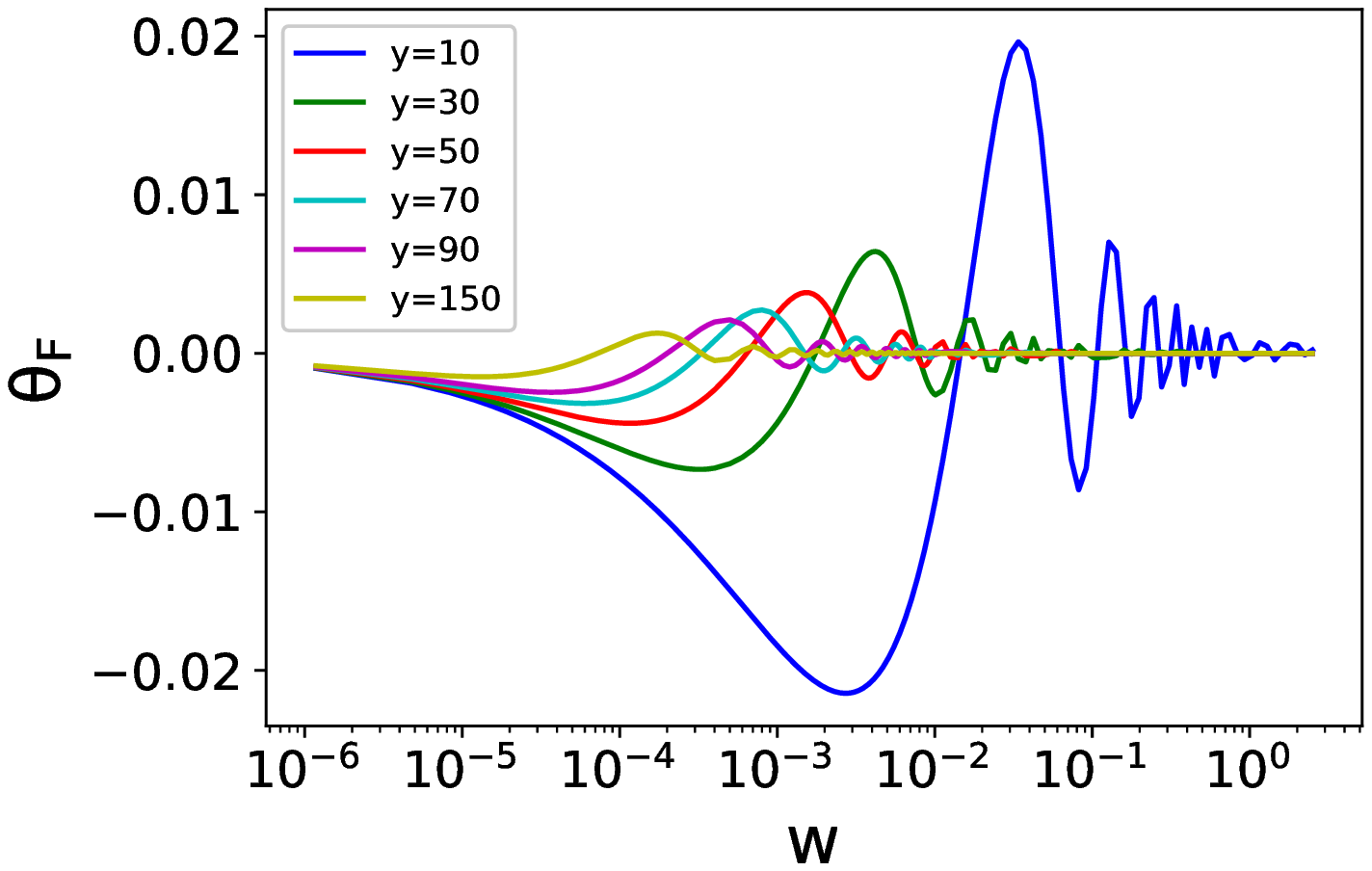}

    \caption{The amplification factor (upper panel) and the phase-change factor (lower panel)
as a function of the dimensionless frequency according to the SIS model. Different lines correspond to
different impact parameters ($y$). Note that this is only a demonstration of some selected amplification 
factors. We also calculate the same quantities for $y = 1,5,20,40,60,80,100$.
}
    \label{FigF}
\end{figure}

Figure~\ref{FigF} shows the dependence of  $|F|$ and $\theta_F$ on the
dimensionless frequency $w$ and impact factor $y$. We notice three results which
are important for the later estimation of the lensing probability. 

First, the amplification factor $|F|$ in general decreases with increasing
impact parameter $y$. However, even when $y$ is relatively large, e.g.,
$y=90$, the amplification factor converges to the value in the geometric
limit and is not $1$, and the phase-change factor does
not vanish. The implication is that even though the wave-optics effect is weak,
it may still be detectable if the SNR of the event is sufficiently large.  We
will study the criterion for detecting such a weak signal in the later sections.  The
previous works, however, normally adopt an upper limit between $y=1$ and $10$ to
estimate the lensing probability \citep[e.g.,][]{takahashi03}. Such a small
value could cause an underestimation of the number of lensing events. 

Second, when $y$ is fixed, both $|F|$ and $\theta_F$ could vary significantly
due to the change of $w$.  In particular, the critical value of $y$, above
which the diffraction effect becomes undetectable, depends on $w$, which,
according to Equation~(\ref{Eq7}), depends on the lens mass, lens redshift, and
the GW frequency.  We will take such a dependence into account in the following
sections. These results indicate that it is oversimplified to use an single
value of $y$ to estimate the lensing probability in the diffraction limit, as
is often the case in the previous works. 

Third, the peaks of the amplification factor and the phase-change angles shift
to smaller $w$ as $y$ increases. As a result, for large impact parameters,
i.e., $y=10-150$, the wave-optics effect appears the most significant at a
small value of frequency, e.g., $w\sim(10^{-5}-10^{-3})$. Such a small dimensionless
frequency corresponds to a low-mass lens according to Equation~(\ref{Eq7}),
which is about
\begin{eqnarray}
M_L(1+z_L)\simeq800\,M_\odot\left(\frac{w}{10^{-4}}\right)\left(\frac{f}{10^{-3}{\rm Hz}}\right)^{-1}.
\label{eq:MLz}
\end{eqnarray}
The corresponding halo mass is also small, about $10^5-10^7M_\odot$ according
to Appendix~\ref{append:A}.  
The above relationships suggest that the majority of the diffraction
events detected by LISA should be induced by small halos, because (i) the 
lensing probability increases with $y^2$ and (ii)
when $y$ is large only small halos produce strong diffraction effect.

\subsection{DM Halos and Subhalos}\label{subsec:DMdis}

The lenses of our interest are those DM halos as small as
$10^5-10^7\,M_\odot$. The last section has shown that they induce 
an observable
diffraction effect to the mHz GWs in the LISA band.  Two types of DM halos
fall in this mass range.

The first type reside in the low-density regions of the universe.  They
predominate the low-mass end of the mass function of ordinary DM halos
\citep[e.g.][]{Wang_2020}. To compute the number density of these halos, we
adopt the Sheth Tormen halo mass function $dn/dM_h$ \citep[see][ for a review]{COORAY_2002},
where $n$ denotes the number density of halos in unit of ${\rm Mpc}^{-3}$ and
$M_h$ is the halo mass. Note that by convention $dn/dM_h$ has a unit of $\rm
M_\odot^{-1}\, Mpc^{-3}\,h^3$. 

The second type of DM halos fall in our interested mass range are the
substructures of those massive DM halos. These substructures are often
referred to as ``subhalos''.  Numerical simulations show that given the mass
$M_h$ of a main halo, the masses of the subhalos follow a power-law
distribution with a universal power-law index
\citep[e.g.][]{Gao_2004a,Gao_2004b,Diemand_2004,Libeskind_2005,Giocoli_2008}).
Following \citet{Han_2016}, we write the mass function of the subhalos as
\begin{eqnarray}\label{Eq12}
    \frac{dN(<R)}{d\ln m}=A(R)\frac{M(<R)}{m_0}\left[\frac{m}{m_0}\right]^{-\alpha},
\end{eqnarray}
where $N(<R)$ is the number of subhalos within a radius of $R$ of the main
halo, $m$ is the mass of the subhalo, $A(R)$ is a normalization factor,
$M(<R)$ is the total mass enclosed by the radius $R$, and $\rm
m_0=10^{10}\,M_\odot$ and $\alpha=0.96$ are constants nearly independent of
the halo mass or redshift.  
In the later calculation, we are mainly interested in the
number of subhalos within the virial radius $R_{\rm vir}$ of the main halo,
regardless of their spatial distribution within the main halo.
Therefore, we should replace $R$ with $R_{\rm vir}$ when using 
Equation~(\ref{Eq12}).  By construction, the total mass within the virial
radius is $M(<R_{\rm vir})=M_h$. This leaves $A(R_{\rm vir})$ the only
quantity that is undetermined. We notice that Figure~15 of \citet{Han_2016}
gives the value of $A(<R)$ which shows that at $R=R_{\rm vir}$ the value
converges to $0.01$ for a wide range of halo mass, from $M_h= 10^{12}\,
h^{-1}M_\odot$ to $10^{15}\, h^{-1}M_\odot$. For this reason, we adopt
$A(<R_{\rm vir})=0.01$ for our later calculations.

Knowing the mass function of subhalos in one main halo, we can calculate
the mass function density at a given redshift for all the subhalos of the 
same mass with
\begin{eqnarray}\label{Eq13}
	\frac{dn_{\rm sub}}{dm}(m,z_L)=\int dM'_h \frac{dn(M'_h,z_L)}{dM'_h}\frac{dN}{d\ln m}\frac{1}{m}.
\rm \end{eqnarray}
Such a quantity is useful for our later calculation of the lensing
probability. Correspondingly, 
the total mass function density contributed by the
halos and subhalos of a mass of $M_h$ is
\begin{eqnarray}\label{eq:lensMFD}
	\xi_{\rm lens}(M_h, z) = \frac{dn_{\rm sub}}{dM_h}(M_h,z)+\frac{dn}{dM_h}(M_h,z).
\end{eqnarray}
Note that to find the lens mass $M_L$ corresponding to a halo mass $M_h$,
the relationship derived in Appendix~\ref{append:A} is applied.

Figure~\ref{fig:DMSHMFD} shows the mass function density predicted by
Equation~(\ref{eq:lensMFD}). We can see that the mass function density has
little evolution from redshift $z=2$ to $6$ (upper panel), and it is more
sensitive to the halo mass (lower panel). We note that in general halos are
more numerous than subhalos. Nevertheless, we include subhalos in the
calculation for completeness.

\begin{figure}
    \centering
    \includegraphics[width=0.45\textwidth]{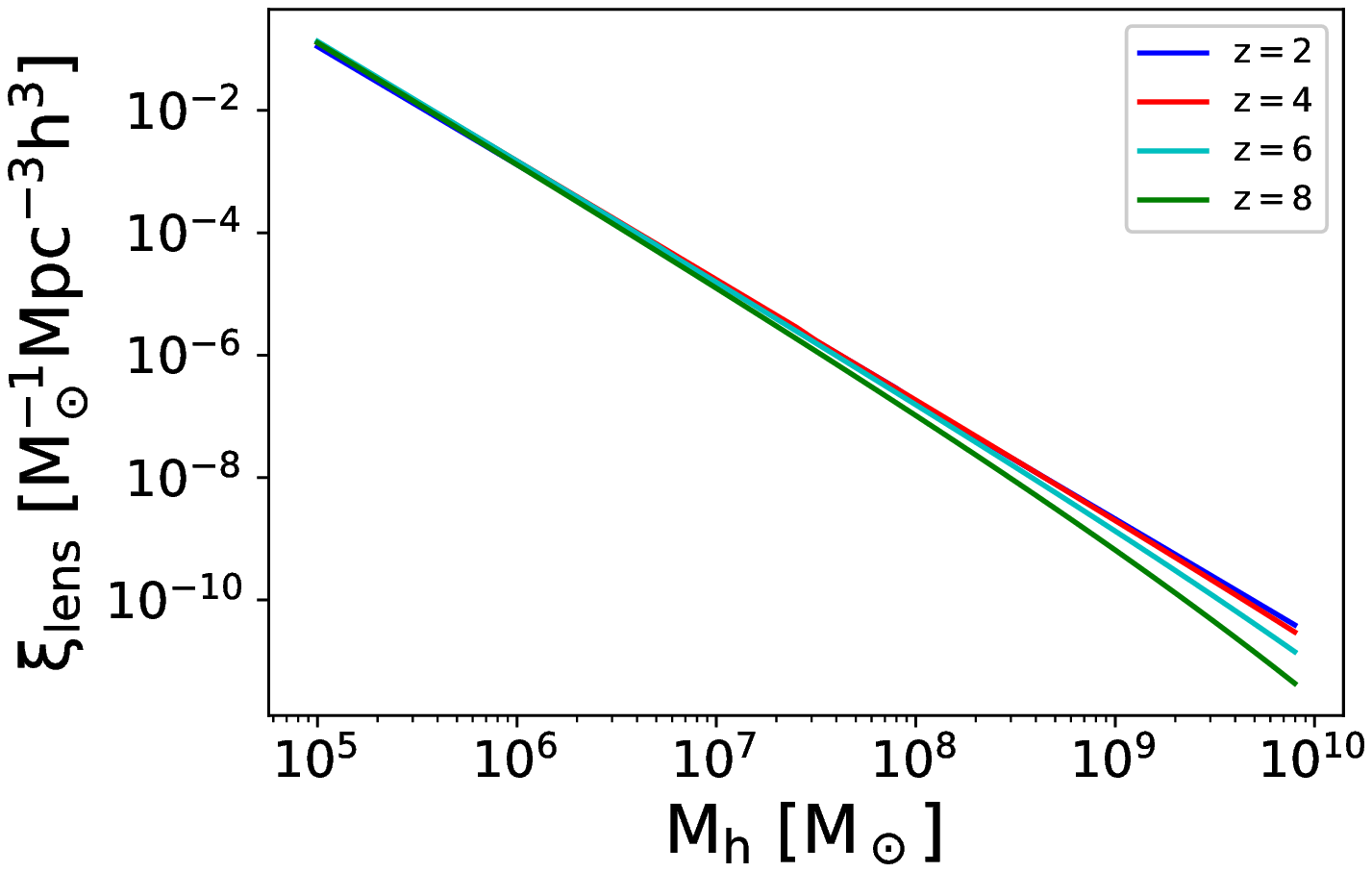}
    \includegraphics[width=0.45\textwidth]{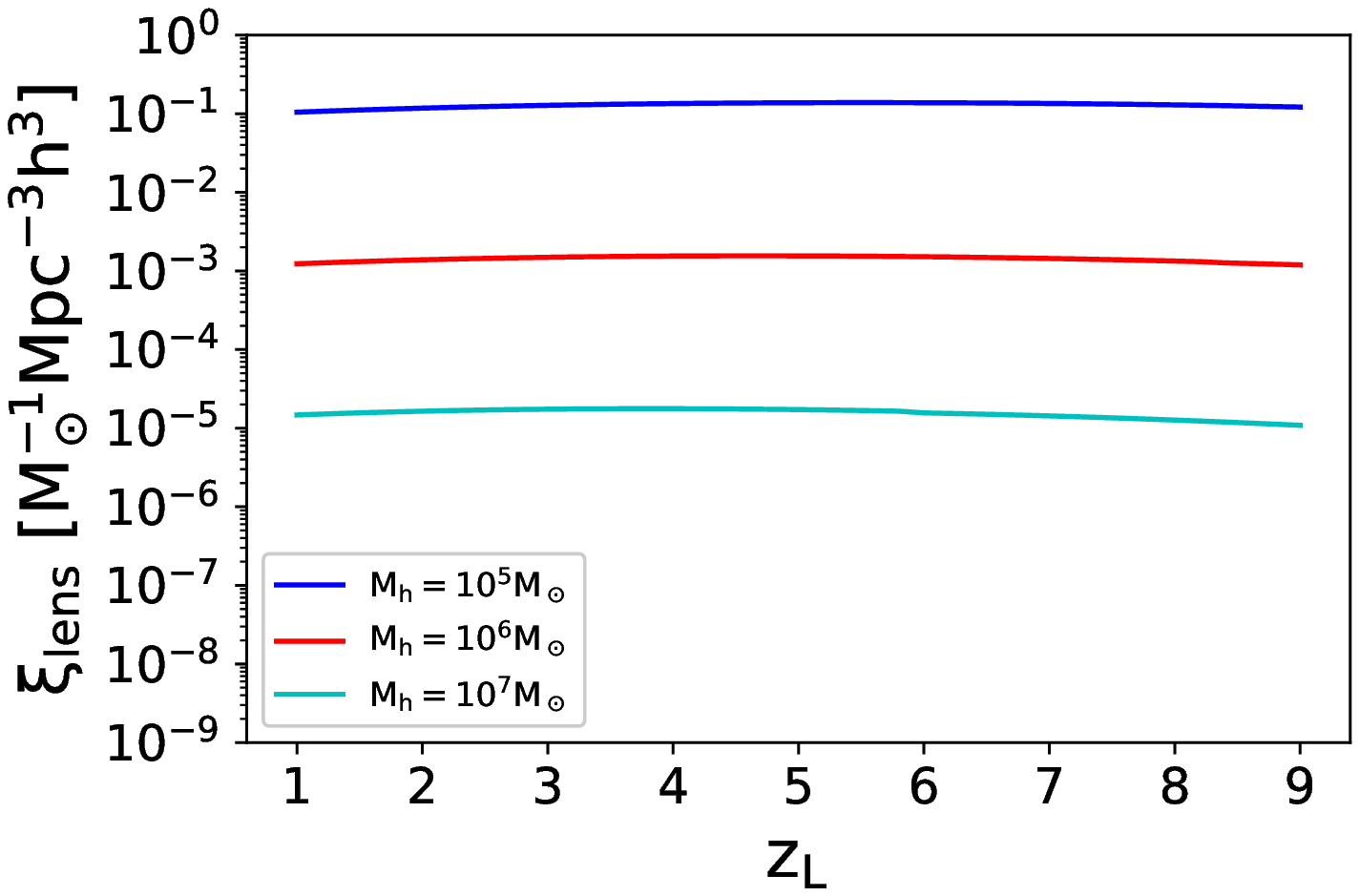}
	\caption{The mass function of the small halos which could produce diffraction effects 
	as a function the halo mass (upper panel) or redshift (lower panel).
}
    \label{fig:DMSHMFD}
\end{figure}

\subsection{Calculation of the Lensing Probability}\label{subsec:EER}

To calculate the lensing probability, we have to specify (i) the number of
lenses of different masses at each redshift and (ii) the solid angle these
lenses cover in which we can detect the diffraction of GW. 

For (i), we start with the halo mass function density derived in the previous
section, $\xi_{\rm lens}(M_h, z)$. Since the SIS model predicts a unique
relationship between the lens mass and halo mass, $M_h(M_L,z_L,z_S)$ (see
Appendix~\ref{append:A}), we can rewrite $\xi_{\rm lens}$ as a function of the
lens mass, i.e., $\xi_{\rm lens}(M_h(M_L,z_L,z_S), z_L)$. Using this new
mass function, we can calculate the number of lens
in the mass range $(M_L,M_L+dM_L)$ and redshift bin $(z_L,z_L+dz_L)$ per unit
solid angle using the equation
\begin{eqnarray}
	\frac{d^3N_{\rm lens}(M_L,z_L,z_S)}{dM_Ldz_Ld\Omega}&=&\xi_{\rm lens}(M_h(M_L,z_L,z_S),\, z_L)\nonumber\\
	&&\times\chi^2(z_L)\frac{d\chi}{dz}\frac{dM_h}{dM_L},\label{eq:Nlens}
\end{eqnarray}
where $\chi(z)$ is the comoving distance for redshift $z$.

\begin{figure}
    \centering
    \includegraphics[width=0.47\textwidth]{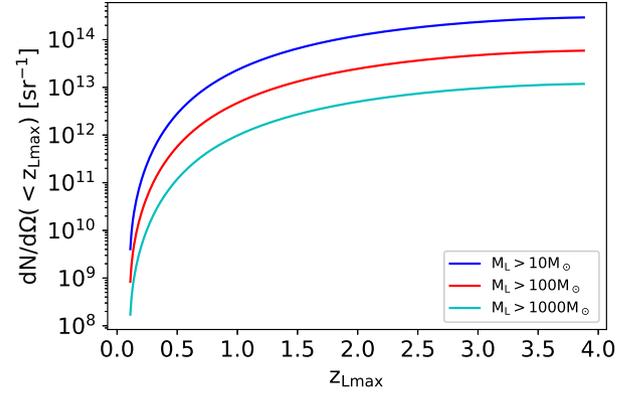}
    \caption{Cumulative distribution of lenses per unit solid angle. The three curves from top to bottom are 
counting the lenses above three different minimum masses, i.e., $M_{\rm Lmin}=(10,\,100,\,1000)\, M_\odot$. 
In this example, the source is at 
$z_S=4$. }
    \label{fig:N_int}
\end{figure}

Figure~\ref{fig:N_int} shows the result of Equation~(\ref{eq:Nlens})
integrated over a redshift range of $[0,z_{\rm Lmax}]$ and above a certain
lens mass.  The source is assumed to be at $z_S=4$.  It is clear that the
number of lenses in a solid angle increases with redshift, and small lenses
(e.g., $M_L\sim10\,M_\odot$) are the most numerous.  Therefore, we expect that
small halos contribute most of the lensing events. We have considered
the lenses as small as $10M_\odot$ because they correspond to a halo mass of about
$10^5M_\odot$ (see Appendix~\ref{append:A}).

As for (ii), suppose $y_{\rm crit}$ is the critical impact parameter in the source plane within which
the effect due to the diffraction of GW is detectable. In the lens plane, the critical impact parameter
corresponds to an angular size of 
\begin{eqnarray}
\theta(M_S,M_L,z_L,z_S)=\xi_0 y_{\rm crit}/D_L.
\end{eqnarray} 
We have written $\theta$ as a function of the source mass $M_S$ and redshift $z_S$ to
highlight the dependence of $y_{\rm crit}$ on the ``loudness'' of the source.
Therefore, the
lensing effect is detectable within a solid angle of
\begin{eqnarray}\label{Eq16}
    \pi\theta^2 = \pi y^2\times \frac{4GM_L D_{LS}}{c^2D_L D_S}
\end{eqnarray}
towards the lens, where we have used the relation
\begin{eqnarray}\label{Eq15}
    \xi_0 = 2\sqrt{GM_L\frac{D_L D_{LS}}{c^2D_S}}
\end{eqnarray}
from the SIS model. Summing up all possible lenses between the source and the observer, 
we derive the lensing probability--the probability of detecting the diffraction effect 
in a given GW source--as
\begin{eqnarray}\label{Eq18}
P 
= \int_0^{z_S} dz_L
\int
\pi\theta^2\times\frac{d^3N_{\rm lens}(M_L,z_L,z_S)}{dM_Ldz_Ld\Omega}
dM_L.
\end{eqnarray}

In principle, the integration should be performed over all possible lens
masses.  In practice, we restrict the integration within a mass range $[M_{\rm
Lmin},M_{\rm Lmax}]$.  The upper and lower limits are functions of lens
redshift $z_L$, which should be determined by evaluating the prominence of the
diffraction effect. Only those lenses producing a detectable diffraction effect
should be counted. The following subsection explains how we quantify the
detectability of the diffraction effect.

\subsection{Signal and Matched Filtering}\label{subsec:matched-filtering}

The magnification factor
derived in Section~\ref{subsec:lensing_model} is a function of GW
frequency. To use it, we need to first derive the unlensed GW signal in the 
frequency domain. This is done by a Fourier transformation,  
\begin{eqnarray}\label{Eq24}
    \Tilde{h}(f)=\int e^{2\pi ift}h(t) dt,
\end{eqnarray}
of the GW strain $h(t)$ in the time domain.

\begin{figure}
    \centering
    \includegraphics[width=0.9\linewidth]{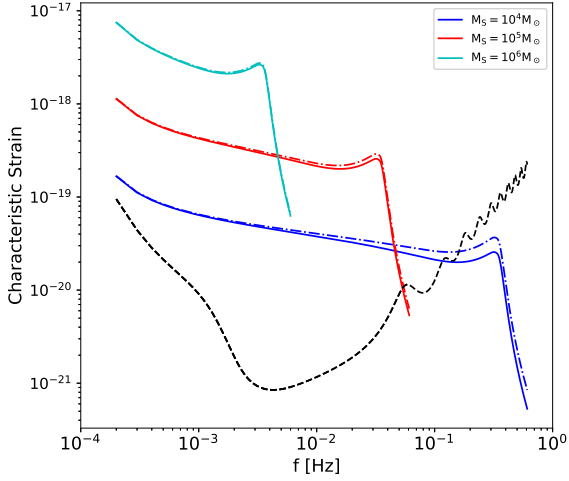}
    \caption{Characteristic strains of unlensed (solid lines) and lensed (dot-dashed lines)
MBH binaries. The lines of different colors correspond to different total masses of the binaries. The black dashed line is the square root of the spectral noise density 
of LISA, i.e., $\sqrt{fS_h(f)}$, adopted from \citet{Robson_2019}.
}
       \label{fig:SourceMass}
\end{figure}

For illustrative purposes, we show in Figure~\ref{fig:SourceMass} the
characteristic strain of three MBH mergers (solid curves).  We assume
equal-mass mergers with zero eccentricity, zero spins and zero inclination for
simplicity and the total masses are $M_S=10^4$, $10^5$, and $10^6\, M_\odot$
respectively. The source redshift is fixed at $z_S=4$ in these examples.  The
waveforms are generated using the ``IMRPhenomC'' model in the PyCBC package
\citep{santamaria20,alex_nitz_2020_4075326}, excluding the effect of BH
spin. As is mentioned in \citet{2018PhRvD..98j4029D}, spin and
eccentricity could also introduce diffraction-like waveform modulation.
However, they also found that
for misaligned
spin, the induced pattern is more densely packed at low frequencies while the diffraction
effect is spread across the frequency domain. The amplitude modulation due to spin is
also much higher than its phase modulation, which is distinctive from the
diffraction effect. As for eccentricity, it induces high harmonics,
which is a feature absence from diffraction effect. Based on these differences, 
we assume that the modulation of the waveform by
spin and eccentricity can be modeled in the future. Here
we focus only on the diffraction effect. 
For the following calculation of unlensed
template, we only vary the $M_S$ and $z_S$ parameters. Because we assume
circular orbits for the MBH binaries, the merger time is about $1.7f_{\rm
mHz}^{-8/3}M_4^{-5/3}(1+z_S)$ years \citep{peters63}, where $f_{\rm mHz}$ is
the GW frequency in unit of mHz and $M_4$ is the total BH mass in unit of
$10^4M_\odot$. It is shorter than the canonical lifetime of LISA ($5$ years)
except in the case of the smallest BHs. 

We now integrate the characteristic strain in Figure~\ref{fig:SourceMass}
to derive the SNR of each merger.
The calculation takes advantage of an inner product \citep{Finn_1992,Cutler:1994ys} 
which is defined as 
\begin{eqnarray}\label{Eq20}
    \langle h_1|h_2\rangle = 2\int_0^\infty \frac{\Tilde{h}_1^*(f)\Tilde{h}_2(f)+\Tilde{h}_1(f)\Tilde{h}_2^*(f)}{S_h(f)}df,
\end{eqnarray}
where $h_1$ and $h_2$ are two waveforms, $S_h(f)$ is the one-sided power spectral density for LISA \citep[from][]{Robson_2019}, and the star symbols denote the
complex conjugates.  The SNR of a signal $h$ is defined as ${\rm
SNR}:=\sqrt{\langle h|h\rangle}$.

\begin{figure}
    \centering
    \includegraphics[width=0.45\textwidth]{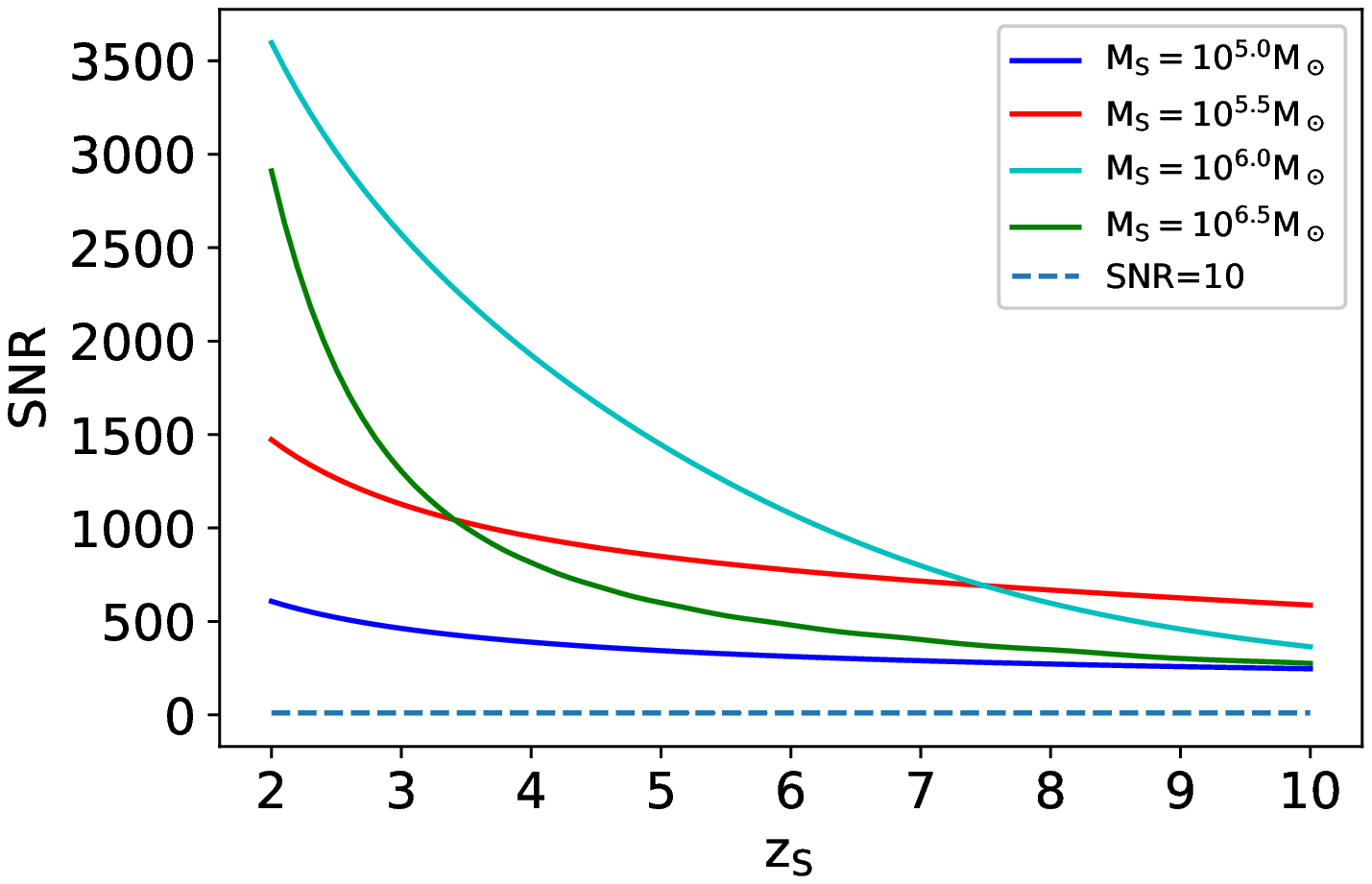}
    \includegraphics[width=0.45\textwidth]{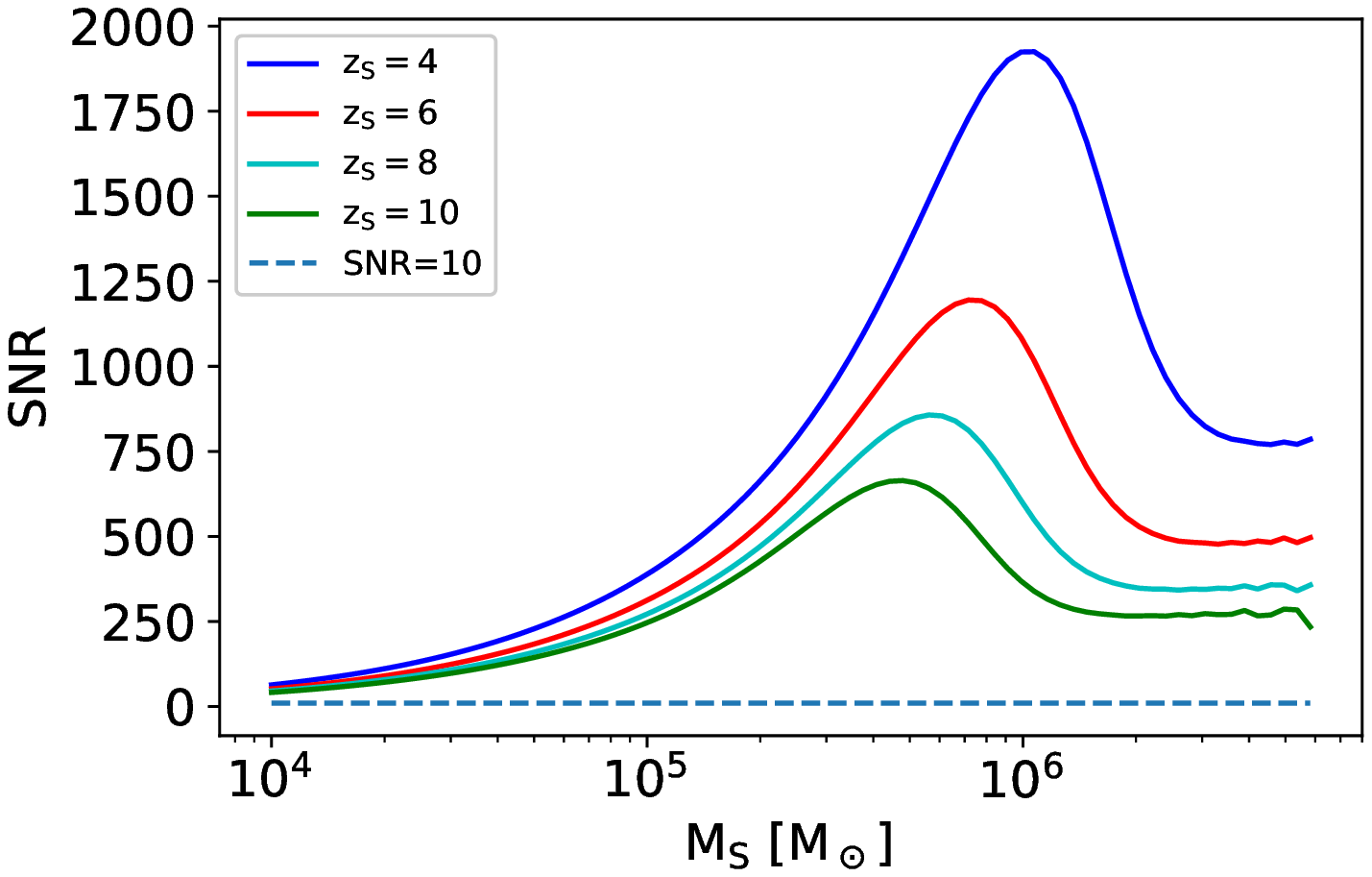}
    \caption{SNR as a function of the redshift (upper panel) 
or the total mass (lower panel) of the source. 
}
    \label{fig:SNR}
\end{figure}

Figure~\ref{fig:SNR} shows the SNR of different MBH mergers at different redshift.
We see that when the total mass is higher than about $10^4M_\odot$ and the source redshift
is lower than $10$, the source in general has a SNR much higher than $10$. These events
are detectable by LISA \citep{Amaro-Seoane_2017}. In the following we study the
lensing signals of these events.

The strain of the lensed signals are shown in Figure~\ref{fig:SourceMass} as the
dot-dashed lines. In the calculation, we assumed that the lens has a mass of
$M_L=10^4\,M_\odot$ and is at a redshift of $z_L=2$. The impact parameter is
set to $y=1$ to maximize the effect in these examples. In this case, we can
discern by eye that the lensed signals differ from the unlensed ones. 

In more general cases, the impact parameters are much larger than $1$ so that
the diffraction effects are much more difficult to discern by eye (e.g., see
Figure~\ref{FigF}). Therefore, we employed the matched-filtering
technique to quantify the deviation of a lensed signal from
a waveform in the template bank.
Suppose $h_1$ is the lensed signal and $h_2$ is an unlensed template,
the difference $\delta h:=h_1-h_2$ is discernible when
the SNR of the difference is larger than 1, i.e., $\inp>1$ \citep[see][for a proof]{lindblom08}.

In our work, $h_2$, the unlensed waveform, is generated from the aforementioned PyCBC package.
We explore the parameter space of the template bank until we find the minimum value of $\inp$.
If this minimum $\inp$ is still greater than $1$, we deem the lensing signal detected.
We call the corresponding waveform the ''best fit''. Note that the best-fit MBH binaries
may differ from the real ones because of lensing.

We are being optimistic in adopting this criterion because we assume that
the deviation of the waveform from the theoretical waveform comes completely
from the lensing effect. However, when the SNR is high, the criterion can be
satisfied due to other factors, such as the presence of other signals, 
inaccurate waveform template, and the
non-Gaussianity/non-stationarity of the noise. 
Nevertheless, it provides a practical criterion by which we can select from our
simulations the lensing events which contain possibly discernible diffraction
features.

\section{Difference between the lensed and the best-fit waveforms}
\label{sec:diff}

The significance of the diffraction effect on the lensing signal depends on
five parameters. Two of them are related to the source, i.e., the total mass
of the MBH binary $M_S$ and the source redshift $z_S$. Two are related to the lens,
i.e., the lens mass $M_L$ and redshift $z_L$. The final one is the impact parameter
$y$. In this section, 
we choose a grid of typical values for these five parameters and  
we investigate how the variation of their values affects 
$\inp$. More specifically, we choose
$M_S = (10^{5.0}, 10^{5.5}, 10^{6.0}, 10^{6.5})\ M_{\odot}$; 
$M_L=(16, 80, 160, 800, 1600, 8000, 16000)\ M_{\odot}$;
$z_S = (4, 6, 8, 10)$. 
The value of $z_L$ depends on $z_S$ and we 
choose $z_L=(0, 1/5, 2/5, 3/5, 4/5)z_S$ unless mentioned otherwise.

\begin{figure}
\includegraphics[width=0.45\textwidth]{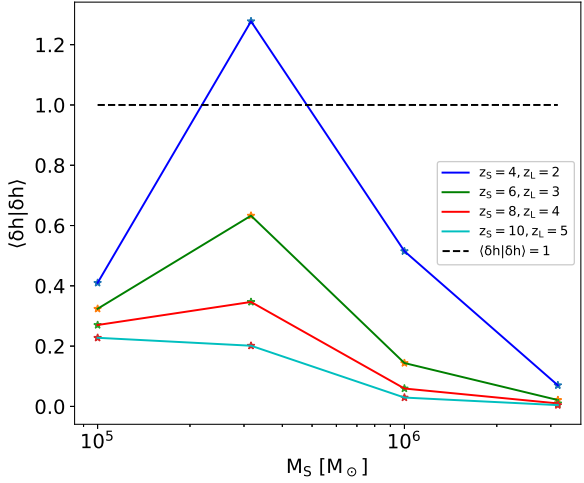}
\caption{The inner product $\inp$ as a function of the total mass of the source.
The solid lines of difference colors correspond to different combinations of source redshift,
and lens redshift. In this example,  
the lens mass is set to $M_L=16\rm \,  M_\odot$
and the impact parameter to $y=40$.
The black dashed line marks the place where $\inp=1$. }
\label{fig:MS}
\end{figure}

Figure~\ref{fig:MS} shows the dependence of $\inp$ on the source mass $M_S$.
Comparing it with the lower panel of Figure~\ref{fig:SNR}, we find that the
inner product behaves similarly as the SNR of the unlensed signal.  The
reason is that higher SNR normally makes the deviation between the lensed signal
and the best-fit template more discernible.

\begin{figure}
    \includegraphics[width=0.45\textwidth]{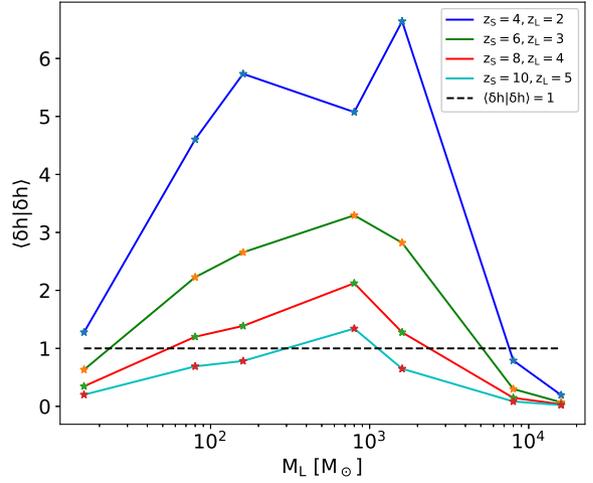}
    \caption{The same as Fig.~\ref{fig:MS} but varying the lens mass $M_L$ while fixing the source mass
at $M_S=10^{5.5}\rm\, M_\odot$. The impact parameter is fixed at $y=40$ for easier comparison.}
    \label{fig:ML}
\end{figure}

The dependence of $\inp$ on the lens mass $M_L$ is shown in
Figure~\ref{fig:ML}.  In this example with a high impact parameter $y=40$, we
find that  $\inp$ first increase and then decrease with the lens mass.
The peak corresponds to a lens whose Einstein radius is
comparable to the wavelength of the GW.  We can also understand the result
through the dimensionless frequency $w$ and the corresponding amplification
factor. On one hand, when $M_L$ is small, $w$ is small. According to
Figure~\ref{FigF}, the lensing effect is small. On the other hand, when $M_L$
is particularly large so that $w$ approaches unity, the system enters the
geometric-optics regime where at any frequency the GW is amplified by the same
factor $\sqrt{1+1/y}$. The wave-optics effect, which is
a frequency-dependent amplification of GWs, diminishes in this case.

\begin{figure}
    \centering
    \includegraphics[width=0.45\textwidth]{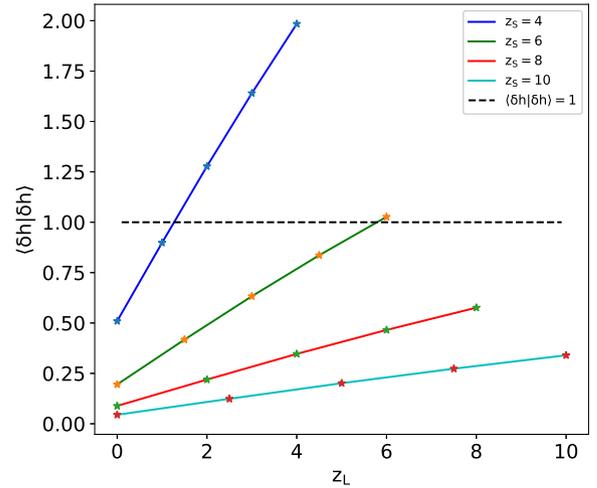}
    \caption{Dependence of the inner produce $\inp$ on the lens redshift $z_L$. In this example,
the source mass is chosen to be $M_S=10^{5.5}\rm\, M_\odot$ and the lens mass is $M_L=16\rm\, M_\odot$. 
The impact parameter is fixed at $y=40$ for easier comparison.}
    \label{fig:zS_zL_y}
\end{figure}

Figure~\ref{fig:zS_zL_y} shows the inner product as a function of the lens
redshift.  We find that $\inp$ increases as the lens redshift approaches the
source redshift.  This behavior is caused by the fact that the lensing effect
is in general stronger when the lens and the source are closer. We also find
that the sources at higher redshift in general produce a smaller  $\inp$. This
result stems from the decrease of the SNR as the source redshift increases.
Note that the effect of lens redshift should be similar, but relatively weak, compared to
the effect induced by the
lens mass. This is because they affect $w$ through the product
$(1+z_L)M_L$. While $z_L$ can only vary by a factor of a few, 
$M_L$ can change by orders of magnitude.

\begin{figure} \centering
\includegraphics[width=0.45\textwidth]{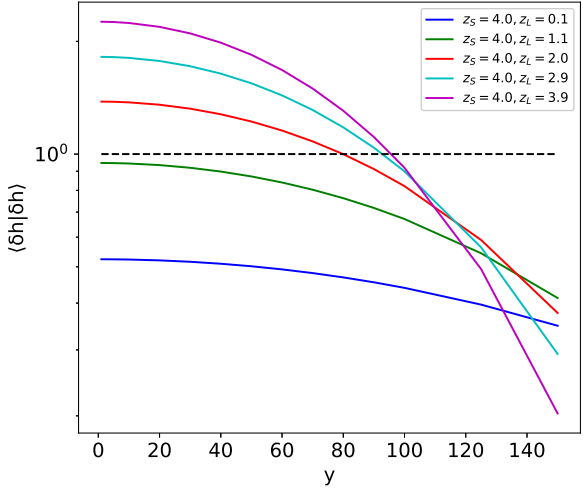} 
\caption{The
inner-product $\inp$ as a function of the dimensionless impact parameter $y$.
These examples assume $M_S=10^{5.5}\rm\, M_\odot$ and $M_L=16\rm\,
M_\odot$.} \label{fig:MF_y} \end{figure}

Finally, we show the dependence of $\inp$ on the impact parameter $y$ 
in Figure~\ref{fig:MF_y}.  
In general, the inner product decreases with the impact
parameter.
Their intersection with the horizontal line $\inp=1$
determines $y_{crit}$. 
We can see that, for some parameters, $y_{crit}$ can be as large as $10^2$.
Such a large impact parameter has not been accounted for
in the previous studies of the wave-optics effect.
It could significantly enhance the lensing probability of LISA MBHs.
 
\section{Lensing Probability}\label{subsec:LP}

Having investigated the dependence of the inner product $\inp$ on the parameters
$M_S$, $z_S$, $M_L$, $z_L$, and $y$, we can now include the halo mass function
and calculate the probability that a MBH binary in the LISA band has $\inp>1$
due to the diffraction effect. We denote this probability as $P(\inp>1)$, and
the expression can be derived from  Equation~(\ref{Eq18}). 

\begin{figure} \centering
	\includegraphics[width=0.45\textwidth]{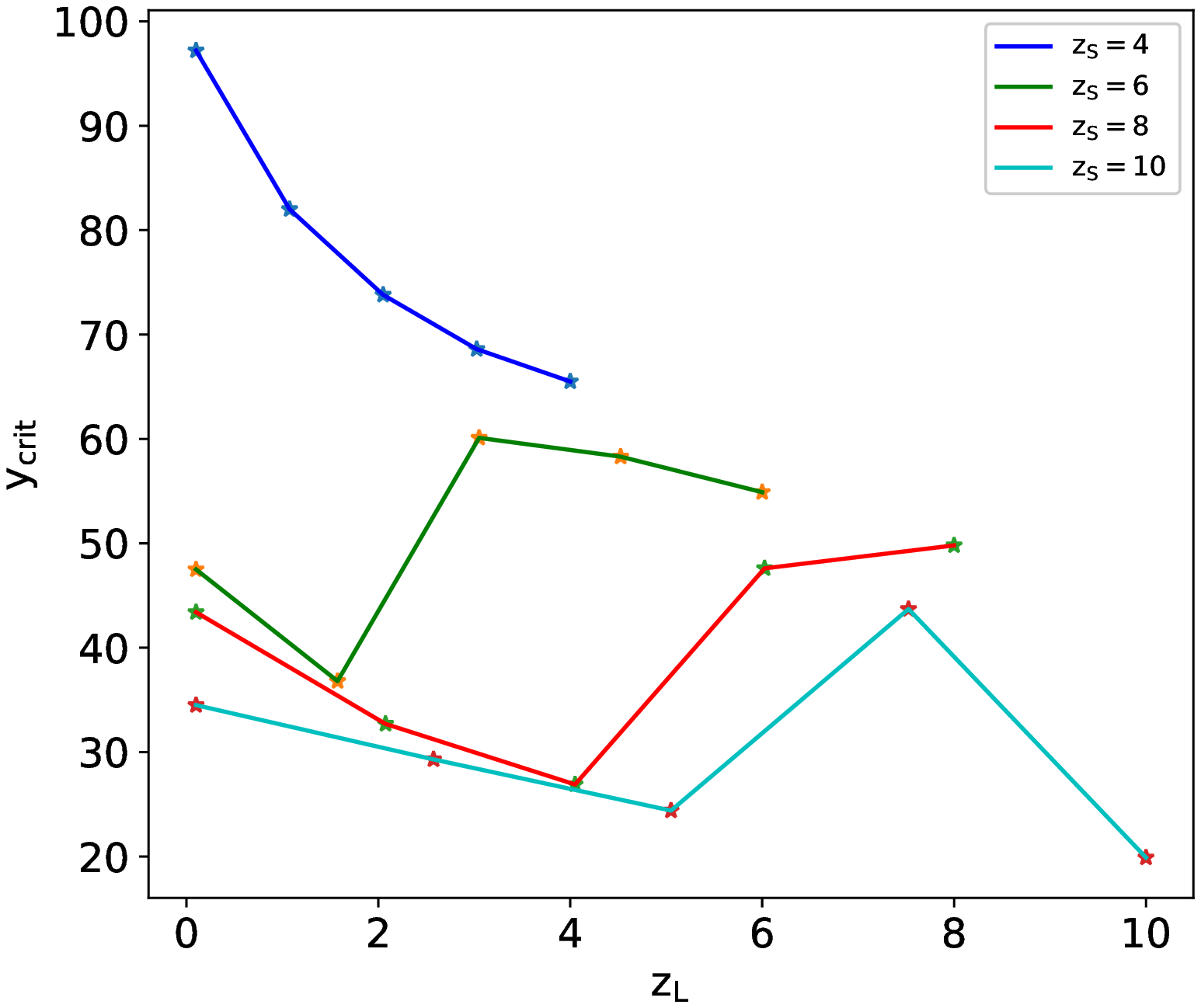}
	\includegraphics[width=0.45\textwidth]{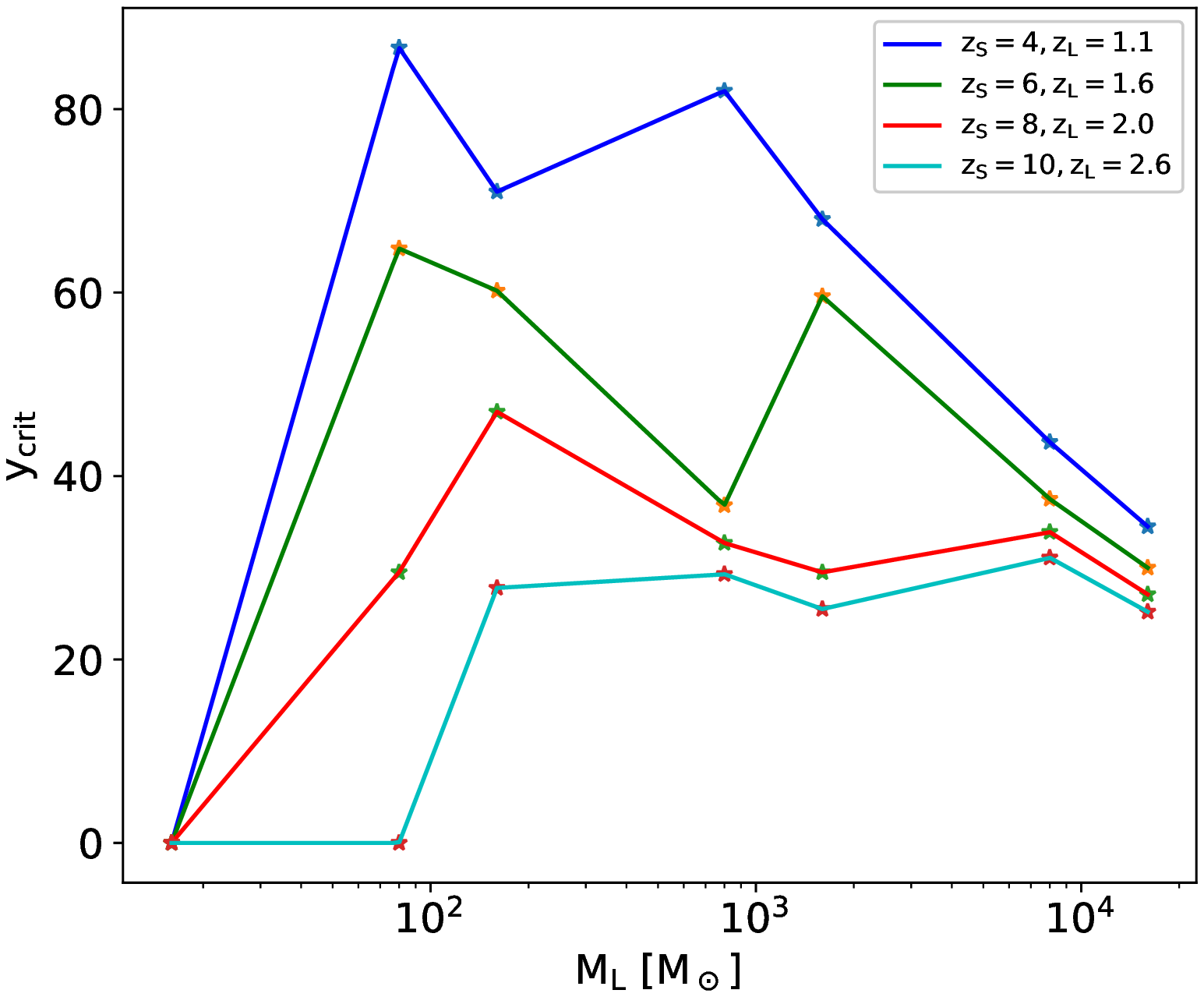}
	\caption{The critical impact parameter which gives $\inp=1$ as a function of 
the lens redshift (upper panel) or lens mass (lower panel). Different lines correspond to different source redshift. In the upper panel, we fixed $M_S = 10^{5.5}\rm\, M_\odot$ and $M_L = 800\rm\, M_\odot$, while in the lower one we fixed $M_S = 10^{5.5}\rm\, M_\odot$. } \label{fig:y_crit} 
\end{figure}

Given $(M_S, \,z_S, \,M_L, \,z_L)$, we first calculate the
critical impact parameter $y_{\rm crit}$ which produces exactly $\inp=1$.  
Figure~\ref{fig:y_crit} shows the critical impact parameter as a function of the
lens redshift and lens mass. In these examples, we have chosen $M_S = 10^{5.5}\rm\, M_\odot$ and varied $z_S$. 
We can see that in a large redshift range, $y_{\rm crit}$ has a value around
40, much higher than the value of $y_{\rm crit}=3$ as has been chosen by the
previous studies \citep[e.g.][]{takahashi03}.  Moreover, sources at lower
redshifts have higher $y_{\rm crit}$, since the higher SNR makes it easier to
discern the wave-optics effect. 
In this work we have calculated the $y_{crit}$ for a total number of
$4\times 7\times 4\times 5=560$ grid points, covering the four-dimensional parameter space of 
 $(M_S,z_S,M_L,z_L)$. Then
by interpolation, we construct the function  
$y_{\rm crit}(M_S,z_S,M_L,z_L)$ which we will use in the following calculation
of the lensing probability.

To calculate the probability, we replace $\theta$ in the integrand of Equation~(\ref{Eq18}) with
$\theta(y_{\rm crit})$ (given by Equation~\eqref{Eq16}) and integrate to give the probability 
$P(y<y_{\rm crit})$. This probability $P(y<y_{\rm crit})$ is equivalent to $P(\inp>1)$.  
In principle, given the source, i.e., after fixing $M_S$ and $z_S$, the $y_{\rm
crit}$ in the integrand is a function of both $z_L$ and $M_L$. In practice, we
only consider the lenses more massive than $M_L=16M_\odot$ because less massive
lenses in general do not produce a sufficiently large inner produce $\inp$, as
Figure ~\ref{fig:ML} has shown. 

The resulting lensing probabilities are given in
Table~\ref{tab:Plensing}. 
We find that, in general, the probability of detecting the wave-optics effect
is about $(0.1-1.6)\%$. Although low, such a probability
 is one order of magnitude larger than that those found in \citet{takahashi03}. 
The enhancement is influenced by a combination of larger impact parameters and
more numerous lenses, but also the decrease of lens mass, which we will discuss in detail in the next section. We note that
the probabilities $P(M_S, z_S)$ derived here
can be used to estimate the number of MBH mergers which show
wave-optics effects prominent enough to be detectable by LISA, once the merger rate of
MBHs as a function of mass ($M_S$) and redshift ($z_S$) is known.

\begin{table*}
    \centering
    \caption{Total Lensing Probability for Different Source Parameters}
    \centering
    \begin{tabular}{c|c c c c}
    \hline
    \hline
    $P(M_S, z_S)$ & $M_S=10^{5.0}\ M_{\odot}$ & $M_S=10^{5.5}\ M_{\odot}$ & $M_S=10^{6.0}\ M_{\odot}$ & $M_S=10^{6.5}\ M_{\odot}$ \\
    \hline
    $z_S = 4$ & 0.0038 & 0.012 & 0.016 & 0.0059 \\
    $z_S = 6$ & 0.0050 & 0.014 & 0.0081 & 0.0024 \\
    $z_S = 8$ & 0.0059 & 0.012 & 0.0056 & 0.00095 \\
    $z_S = 10$ & 0.0057 & 0.0094 & 0.0036 & 0.00040 \\
    \hline
    \end{tabular}

    \label{tab:Plensing}
\end{table*}

In this work, we did not consider a lens mass higher than $1.6\times
10^4\rm\ M_\odot$ because such lenses contribute a small fraction (less than
$10\%$) to the total probability. The reasons are two fold. (i) The
corresponding halos mass is greater than $10^8\rm\ M_\odot$. The number density
of such halos is low.  (ii) As the lens mass exceeds $1.6\times 10^4\rm\
M_\odot$, the lensing effect will approach the geometric limit, diminishing the
detectability of the diffraction effect.

\section{Impact of DM models}\label{sec:DMmodel}

We notice that \citet{takahashi03} derived a lensing
probability of $10^{-4}\sim 10^{-3}$ for the MBH binaries in the LISA band.
It is at least one order of magnitude smaller than our estimation. The discrepancy
stems from the different ranges of lens mass adopted in these two works.

\citet{takahashi03} considered the lenses in the mass range of $M_L =\rm
10^6\sim10^9\, M_\odot$, which corresponds to a halo mass of $10^9\sim
10^{12}\,\rm M_\odot $. Such lenses are already in the geometric-optics
limit. Moreover, they assumed a critical impact parameter of  $y_{\rm crit}=3$.
In our model, we considered the lenses in the mass range of $\rm 1.6\times 10^1\sim
1.6\times 10^4\, M_\odot$. The corresponding halo mass is $\rm 10^5\sim 10^8\,
M_\odot $. These lenses produce wave-optics effect in the lensing signal, and
we have shown that the effect is detectable even for a large impact parameter
of $y\sim40$ (Figure~\ref{fig:y_crit}).

\begin{figure}
    \centering
    \includegraphics[width=0.45\textwidth]{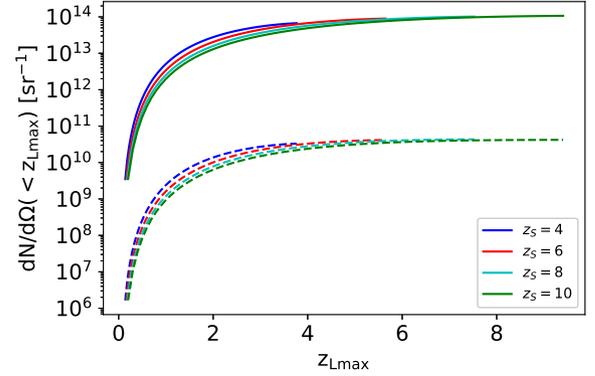}
    \includegraphics[width=0.45\textwidth]{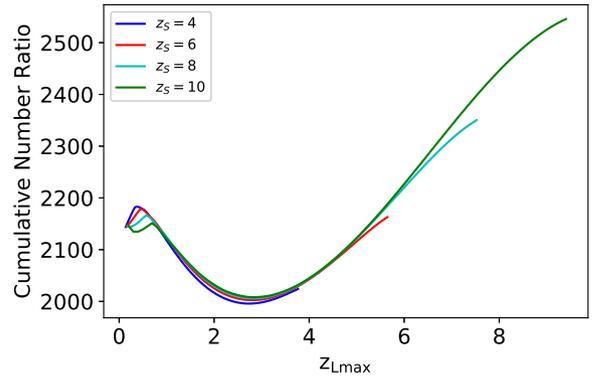}
    \caption{Upper panel: Cumulative distribution of the number of lenses
	as a function of the upper limit of lens redshift $z_{\rm Lmax}$.
The solid lines represent the
	lenses in the mass range of $M_L= 16\sim 1.6\times 10^4\,
	M_\odot$, while the dashed one represent the lenses of $\rm 10^6\sim10^9\, M_\odot$. The four colors refer to
	four different source redshifts. 
	Lower panel: The ratio between the
	two kinds of cumulative lens numbers.} \label{fig:number_difference}
\end{figure}

The difference of the lens masses affects the lensing probability in two ways. (i) In our model, the solid angle within in which
the lensing signal is detectable is about $100$ times smaller than the choice of
\citet{takahashi03} (assuming $y_{\rm crit}=40$), 
since it is proportional to $y_{\rm crit}^2 M_L$
according to Equation~(\ref{Eq16}). (ii) Our lenses are about $1000$ times
more numerous than those considered in \citet{takahashi03} according to
Figure~\ref{fig:DMSHMFD}. The difference can be more clearly seen in
Figure~\ref{fig:number_difference}.  Combining these two consequences, we 
find that our lensing probability is about $10$ times higher
than that derived in \citet{takahashi03}. The case of higher mass source at $z_S=10$ is somewhat different, 
because the SNR reduces significantly. Consequently, $y_{\rm crit}$ 
decreases, so that the probability is only 
several times larger than what \citet{takahashi03} has derived. 

\begin{figure}
    \centering
    \includegraphics[width=0.45\textwidth]{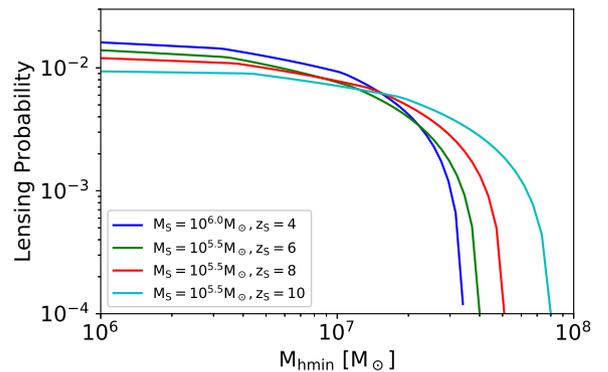}
    \caption{The total lensing probability as a function of the minimum
	halo mass. Different colors refer to difference source mass and redshift. 
}
    \label{fig:LP_masscut}
\end{figure}

The above comparison suggests that the lensing probability is sensitive to the
abundance of small halos. Since different DM models predict very different
number density for small halos, we now investigate the dependence of the
lensing probability on the lower boundary of the halo mass function.  To
simulate the effect of different DM models, we cut off the integration of
Equation~(\ref{Eq18}) at different lower boundaries $M_{\rm hmin}$ and count
only those halo with $M_h>M_{\rm hmin}$. The result is shown in
Figure~\ref{fig:LP_masscut}. We see a sharp cut off around $\rm M_{hmin}=10^8\,
M_\odot$.  Compared to \citet{takahashi03}, our result indicates that the probability of 
detecting the diffraction effect of a MBH binary in the LISA band is significantly enhanced,
because of the numerous small halos in the CDM paradigm. 
It also implies that if warm DM predominates \citep[e.g.][]{Lovell_2014}, the probability of detecting the wave-optics effect
would be low.

\section{Summary Conclusion}\label{sec:conclusion}

In this work, we studied the lensing signals of the MBH binaries in the LISA
band.  We focused on the wave-optics effect and found that it is produced
mainly by the DM halos and subhalos in the mass range of $10^5\sim10^8M_\odot$.
Using the matched-filtering technique, we showed that the effect could be
discernible by LISA even when the source has an impact parameter as large as
$y=40$, or even $y=100$ in some cases. Such a large impact parameter
substantially enhances the probability of detecting the diffraction signatures.
Despite the large impact parameter, the probability of an event lensed
by multiple halos is still low because the Einstein radius is small, normally
$10^3 - 10^4$ times smaller than the virial radius of the host halo.  

According to our preliminary estimation, if CDM predominates the
matter content of the universe, the chance of detecting diffraction effect by
LISA is about more than $1\%$, regarding of the source MBH binaries within the
mass range of $\rm 10^{5.0}\sim10^{6.5}\, M_\odot$. If, on the other hand, warm
DM predominates, the chance of detecting the diffraction effect would be
diminished by at least one order of magnitude.  Therefore, looking for the
wave-optics effects in LISA events could help us constrain the DM models. 

As a final remark, we note that our model of the lensing signal and the
criterion of discerning the diffraction effect are based on ideal assumptions.
For example, we do not consider multiple lenses along the line of sight even
though we have found a relatively high lensing probability.
Moreover, we have assumed that the deviation of the detected signal from the
model template is solely due to gravitational lensing, while for real LISA observation, other
factors, such as confusion between multiple events or inaccuracy of the waveform template, could also contribute to the
deviation. We will address these caveats in a future work.  

\section*{Acknowledgements}
This work is supported by the National Science Foundation of China grants No
11873022, 11991053, and 11805286. X.C. is partly supported by the Strategic Priority
Research Program of the Chinese Academy of Sciences, Grant No. XDB23040100 and
No. XDB23010200. 
Y.H. is partly supported by the National Key Research and Development Program of China (No. 2020YFC2201400), and Guangdong Major Project of Basic and Applied Basic Research (Grant No. 2019B030302001).
We especially thank Liang Dai at University of California,
Berkeley for many insightful discussions on the theory of wave-opitcal lensing
model and precious comments on an early version of this manuscript. We also
extremely thank Xiao Guo at National Astronomical Observatories of China (NAOC)
for providing us with valuable suggestions on various problems encountered in
this work. 

\section*{Data Availability}

The data underlying this article will be shared on reasonable request to the corresponding author.



\bibliographystyle{mnras}
\bibliography{sample63, mybib} 



\appendix

\section{Relation between halo mass and lens mass in the SIS model}\label{append:A}

We assume that DM halo follows an SIS density profile, 
\begin{eqnarray}\label{eq:A1}
    \rho(r) = \frac{\sigma_v^2}{2\pi G r^2},
\end{eqnarray}
where $\sigma_v$ is the velocity dispersion. According to this profile, the
total mass of the halo, $M_h$, is related to the virial radius $r_{\rm
vir}(M_h, z_h)$, which is a function of the mass $M_h$ and redshift $z_h$, as
\begin{eqnarray}\label{halomass}
    M_h =  \frac{2\sigma_v^2}{G}r_{\rm vir}(M_h, z_h).
\end{eqnarray}

The lens mass is defined as the mass enclosed by a circle on the lensing plane with a
radius of $\xi_0 = 4\pi \sigma_v^2 D_L D_{LS}/c^2D_S $, which is known as the Einstein
radius. To calculate the lens mass $M_L$, we use the
the surface density of an SIS projected on the lensing plane, 
$\Sigma(\xi)=\sigma_v^2/(2G\xi)$ \citep{takahashi03}, and derive  
\begin{eqnarray}\label{lensmass}
    M_L = \frac{4\pi^2\sigma_v^4 D_{LS} D_{L}}{GD_{S}c^2}.
\end{eqnarray}
    
\begin{figure}
    \centering
    \includegraphics[width=0.45\textwidth]{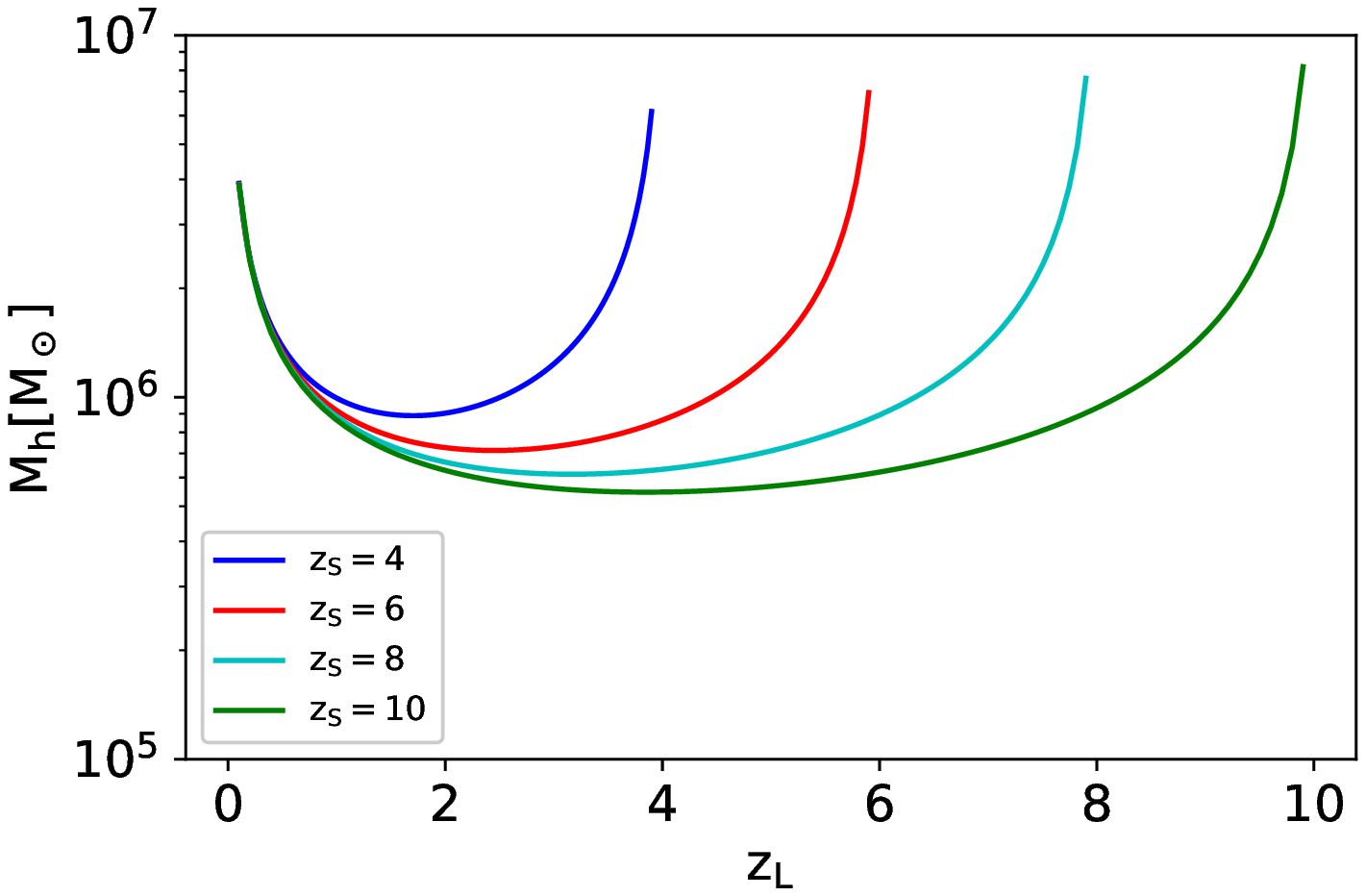}
    \includegraphics[width=0.45\textwidth]{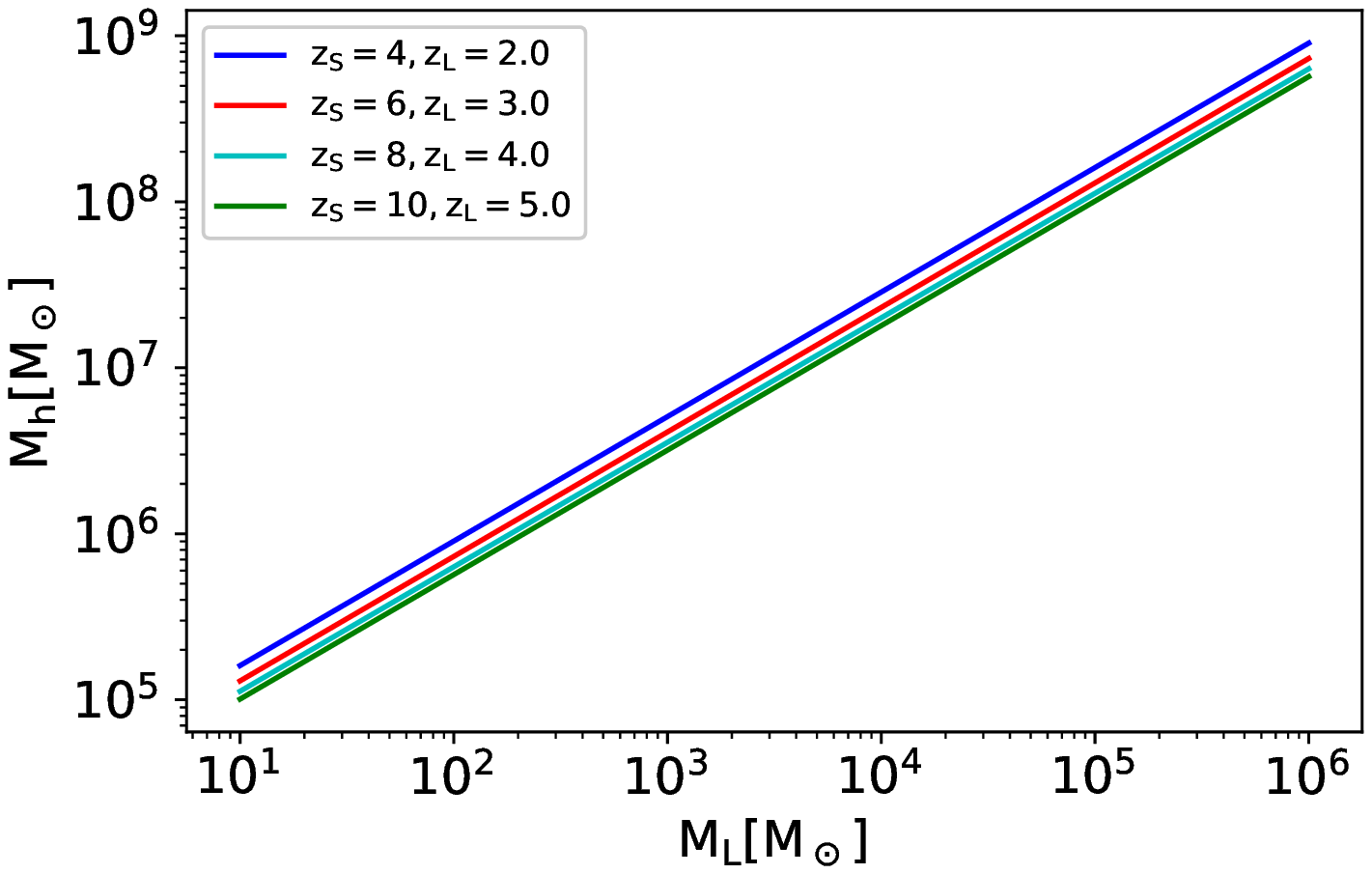}
    \caption{The upper panel shows the halo mass as a function of the redshift
when the lens mass is fixed to $100\,M_\odot$.
Different lines correspond to different redshift of the source.  The lower
panel shows the halo mass as a function of lens mass $M_L$.  Different lines
correspond to different combinations of the source and lens redshifts.
}
    \label{fig:MhML}
\end{figure}

To relate the halo mass to the lens mass, we use Equations~(\ref{halomass}) and 
(\ref{lensmass}) to eliminate $\sigma_v$ and we find that
\begin{eqnarray}\label{A4}
    M_h = r_{\rm vir}(M_h, z_h)\sqrt{\frac{M_L D_S c^2}{\pi^2 G D_{LS} D_L}}.
\end{eqnarray}
Figure~\ref{fig:MhML} shows the relationship between these two masses and the
dependence on the redshift of the halo. 

\bsp	
\label{lastpage}
\end{document}